\documentclass[showpacs,preprintnumbers,amsmath,amssymb]{revtex4}
\usepackage{graphicx}
\begin{document}
%\baselineskip 20.75pt
 %=========================================================================

%\vspace{0.5cm} \preprint{{\em Submitted to Phys.~Rev.~Lett.} %\hspace{3.0in} Web galley}
\date{\today}
\vspace{2.7in}

\title{Bose-Einstein condensation of quasiparticles in graphene}

\author{Oleg L. Berman$^{1}$, Roman Ya. Kezerashvili$^{1,2}$, and Yurii E.
Lozovik$^{3}$} \affiliation{\mbox{$^{1}$Physics Department, New York
City College of Technology, The
City University of New York,} \\
Brooklyn, NY 11201, USA \\
\mbox{$^{2}$The Graduate School and University Center, The
City University of New York,} \\
New York, NY 10016, USA \\
\mbox{$^{3}$Institute of Spectroscopy, Russian Academy of
Sciences,} \\
142190 Troitsk, Moscow Region, Russia}

\begin{abstract}

The collective properties of different quasiparticles in various graphene based structures in high magnetic field have been studied. We predict Bose-Einstein condensation (BEC) and superfluidity of 2D spatially indirect magnetoexcitons in two-layer graphene. The superfluid density and the temperature of the Kosterlitz-Thouless phase transition are shown to be increasing functions of the excitonic density but decreasing functions of magnetic field and the interlayer separation. The instability of the ground state of the interacting 2D indirect magnetoexcitons in a slab of superlattice with alternating electron and hole graphene layers (GLs) is established. The stable system of indirect 2D magnetobiexcitons, consisting of pair of indirect excitons with opposite dipole moments, is considered in graphene superlattice. The superfluid density and the temperature of the Kosterlitz-Thouless phase transition for magnetobiexcitons in graphene superlattice are obtained. Besides, the BEC of excitonic polaritons in GL embedded in a semiconductor microcavity in high magnetic field is predicted. While superfluid phase in this magnetoexciton polariton system is absent due to vanishing of magnetoexciton-magnetoexciton interaction in a single layer in the limit of high magnetic field, the critical temperature of BEC formation is calculated. The essential property of magnetoexcitonic systems based on graphene  (in contrast, e.g., to a quantum well) is stronger influence of magnetic field and weaker influence of disorder. Observation of the BEC and superfluidity of 2D quasiparticles in graphene in high magnetic field would be interesting confirmation of the phenomena we have described.

\vspace{0.1cm}
 %\vspace{0.2 in}


%Key words: polaritons, optical microcavities, quantum wells, %Bose-Einstein condensation, superfluidity

\end{abstract}

\maketitle {}

%\newpage
%-----------------------------------------------------------------------
%-------------------------------------------------------------------------
%-------------------------------------------------------------------------

%-------------------------------------------------------------------------
%-------------------------------------------------------------------------

%-------------------------------------------------------------------------
\section{Introduction}
\label{intro}

The production of graphene, a two-dimensional (2D) honeycomb lattice of carbon atoms that form the basic planar structure in graphite, has been achieved recently  \cite{Novoselov1,Zhang1}.
Electronic properties of graphene, caused by unusual properties of the  band
structure, became the object of many recent experimental and
theoretical studies \cite{Novoselov1,Zhang1, Novoselov2,Zhang2,Falko,Katsnelson,Castro_Neto_rmp}.  Graphene is a gapless semiconductor with massless electrons and holes, described as Dirac-fermions \cite{DasSarma}. The various studies of unique electronic properties  of graphene in a magnetic field have been performed recently \cite{Nomura,Jain,Gusynin1,Gusynin2}. The energy spectrum and the wavefunctions of magnetoexcitons, or electron-hole pairs in a magnetic field, in graphene have been calculated in interesting works~\cite{Iyengar,Koinov}.

The 2D electron system was studied in quantum wells (QWs) \cite{Ando_Fowler_Stern}. The  systems of  spatially-indirect excitons (or pairs of electrons and holes spatially separated in different QWs) in the system of  coupled quantum wells (CQWs), with and without  a magnetic field  were studied in Refs.~[\onlinecite{Lozovik,Lerner,Shevchenko,Berman,Poushnov,Ruvinsky,Berman_Tsvetus,Ovchinnikov}].
 The experimental and theoretical interest to  study these systems is  particularly caused by the possibility of the BEC and superfluidity of indirect excitons, which
can manifest  in the CQWs as persistent electrical currents
in each well and also through coherent optical properties and
Josephson phenomena~\cite{Lozovik,Shevchenko,Berman,Poushnov,Berman_Tsvetus,Ovchinnikov}. The outstanding experimental success was achieved now in this field~\cite{Snoke,Butov,Timofeev,Eisenstein}. The electron-hole pair condensation in the graphene-based bilayers have been studied in~[\onlinecite{Sokolik,MacDonald2,Efetov,Joglekar}].

The collective properties of Bose quasiparticles such as excitons, biexcitons, and polaritons in various graphene-based structures in high magnetic field are very interesting in the relevance to the BEC and superfluidity, since the random field in graphene is weaker than in a QW, particularly, because in a QW the random field is generated due to the fluctuations of the width of a QW.  Let us mention that if the interaction of bosons with the random field is stronger, the BEC critical temperature is lower \cite{BLSC}. In this paper we study the superfluidity of magnetoexcitons in bilayer graphene, instability of the system of magnetoexcitons in superlattice formed by many GLs, superfluidity of magnetobiexcitons in graphene superlattice, and BEC of polaritons in GL embedded in optical microcavity in a trap. Let us mention that all these systems of quasiparticles are considered in high magnetic field. The  BEC of magnetoexcitons in graphene structures  can exist at much lower magnetic field than in  QWs, because the distance between electron Landau levels in graphene is much higher than in a QW at the same magnetic field, and, therefore, the lower magnetic field is required in graphene than in a QW to neglect the electron transitions between the Landau levels.

%-------------------------------------------------------------------------
\section{Effective Hamiltonian of magnetoexcitons and photons in microcavity in high magnetic field}
\label{hamilton}

Recently,  Bose coherent effects of 2D exciton
polaritons in a quantum well embedded in an optical microcavity have been the subject of theoretical~\cite{book}.   and experimental~\cite{Dang,Yamamoto,Baumberg} studies.   To obtain polaritons, two mirrors placed opposite each other form a microcavity, and quantum wells are embedded within the cavity at the antinodes of the confined
optical mode. The resonant exciton-photon interaction results in the Rabi splitting of the excitation spectrum.  Two polariton branches appear in the spectrum due to the resonant exciton-photon coupling. The lower polariton (LP) branch of the spectrum has a minimum at zero momentum. The effective mass of the lower polariton is extremely  small, and lies in the range $10^{-5}-10^{-4}$ of the free electron mass. These lower polaritons form a 2D weakly interacting Bose gas. The extremely light mass of these bosonic quasiparticles, which corresponds to
experimentally achievable excitonic  densities, result in a relatively high
critical  temperature for superfluidity, of $100 \
\mathrm{K}$ or even higher. The reason for such a high  critical temperature is that the 2D thermal de Broglie wavelength is inversely proportional to the mass of the quasiparticle.

While at finite temperatures there is no true BEC  in any infinite untrapped
2D system, a true 2D BEC can exist in the presence of a confining
potential~\cite{Bagnato,Nozieres}.
  Recently,  the  polaritons in a harmonic potential trap have  been studied experimentally in a GaAs/AlAs quantum well embedded in a GaAs/AlGaAs microcavity \cite{Balili}. In this trap,
   the exciton energy is shifted using  stress. In this system, evidence for the BEC of polaritons in a QW
has been observed~\cite{science}. The theory of the
BEC and superfluidity of excitonic polaritons in a QW
without magnetic field in a parabolic trap has been developed in
Ref.~[\onlinecite{Berman_L_S}]. The Bose condensation of polaritons is caused by their
bosonic character~\cite{science,Berman_L_S,Kasprzak}.  However, while the exciton polaritons have been studied in a QW, the formation of the polaritons in graphene in high magnetic field have not yet been considered. Moreover, the magnetopolaritons formed as superposition of magnetoexcitons and cavity photons in magnetic field have not yet been studied. We consider a 2D system of polaritons in graphene layers (GLs) embedded in a microcavity from the point of view of the existence of the BEC within it.

\begin{figure}[tbp]
\includegraphics[width = 3.5in]{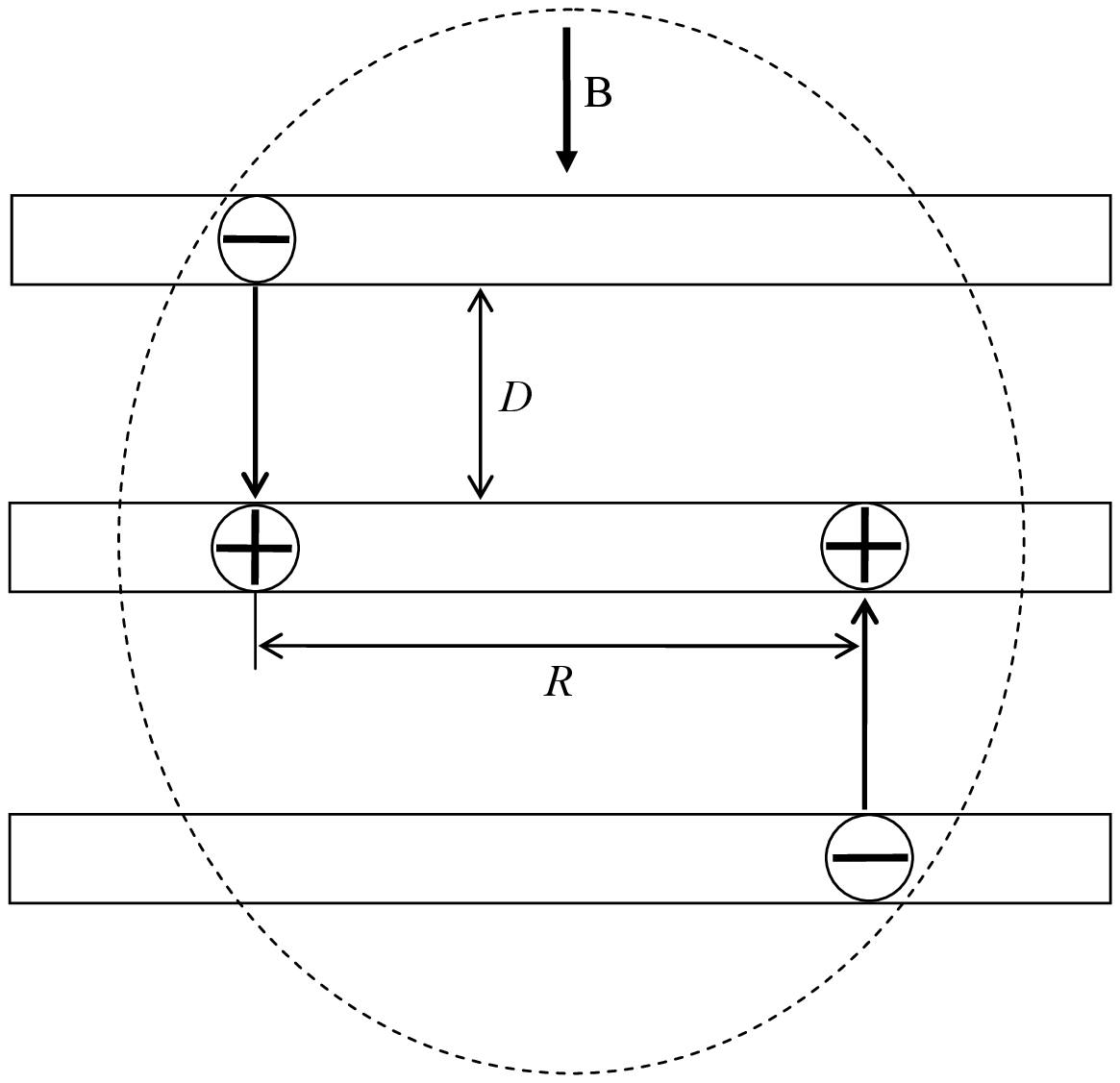}
\caption{Two-dimensional indirect magnetobiexcitons consisting of
indirect magnetoexcitons with opposite dipole moments, located in
neighboring pairs of GLs.} \label{biexciton}
\end{figure}

Lets us consider the most general case when the superlattice with  alternating electronic and hole parallel GLs  in the external field is embedded in an optical microcavity in high magnetic field.
 At the small densities $n$  the system of indirect excitons at low temperatures
is the two-dimensional weakly nonideal Bose gas with normal to wells
dipole moments  $\mathbf{\it{d}}$ in the ground state ($d = eD$, $e$ is the charge of an electron,  $D$
is the interlayer separation). In contrast to ordinary excitons, for the
low-density spatially indirect magnetoexciton system the main
contribution to the energy is originated from the dipole-dipole
interactions $U_{-}$ and $U_{+}$ of magnetoexcitons with opposite
(see Fig.~\ref{biexciton}) and parallel dipoles, respectively. The
potential energy of interaction between two indirect magnetoexcitons
with parallel $U_{+}(R)$ and opposite $U_{-}(R)$ dipoles is a
function of the distance $R$ between indirect magnetoexcitons along
GLs  and is given as
%-------------------------------------------------------------------------
\begin{eqnarray}
\label{inter}U_{+}(R)&=& \frac{2e^{2}}{\epsilon
R}-\frac{2e^{2}}{\epsilon \sqrt{R^{2}+D^{2}}} \ , \nonumber\\
 U_{-}(R)&=&\frac{e^{2}}{\epsilon
R}-\frac{2e^{2}}{\epsilon \sqrt{R^{2}+D^{2}}}+ \frac{e^{2}}{\epsilon
\sqrt{R^{2}+4D^{2}}} \ ,
\end{eqnarray}
%-------------------------------------------------------------------------
where $e$ is the charge of an electron, $\epsilon$ is the dielectric constant.

The Hamiltonian of magnetoexcitons and photons  in the strong magnetic field is
given by
%-------------------------------------------------------------------------
\begin{eqnarray}
\label{Ham_tot_pol} \hat{H}_{tot} = \hat{H}_{mex} + \hat{H}_{ph} +
\hat{H}_{mex-ph} \ ,
\end{eqnarray}
%-------------------------------------------------------------------------
where $\hat{H}_{mex}$ is a magnetoexcitonic Hamiltonian,
 $\hat{H}_{ph}$ is a photonic Hamiltonian, and $\hat{H}_{exc-ph}$ is the
Hamiltonian of magnetoexciton-photon interaction.

Let us analyze each term of the Hamiltonian
(\ref{Ham_tot_pol}). The effective Hamiltonian and the energy dispersion for magnetoexcitons  in graphene layers in a high magnetic field $B$ in the infinite system was derived in Ref.~[\onlinecite{Berman_K_L}].
The  effective Hamiltonian $\hat{H}_{mex}$ of
the low-density system of the indirect magnetoexcitons in high magnetic field  in the
superlattice in the subspace of the lowest Landau level is given by~\cite{Berman_K_L} (we neglect the electrons transitions between different Landau levels due to electron-hole Coulomb attraction)
%-------------------------------------------------------------------------
\begin{eqnarray}
\label{Htotb}
 \hat{H}_{mex}=\hat{H}_{0}+\hat{H}_{int}.
\end{eqnarray}
%-------------------------------------------------------------------------
Here $\hat{H}_{0}$ is the effective Hamiltonian of the system of
noninteracting trapped magnetoexcitons in high magnetic field:
%-------------------------------------------------------------------------
\begin{eqnarray}
\label{h0}
\hat{H}_{0}=\sum_{\mathbf{P}}^{{}}\varepsilon _{mex}(p)(a_{\mathbf{P}}^{+}a_{%
\mathbf{P}}^{{}}+b_{\mathbf{P}}^{+}b_{\mathbf{P}}^{{}}+a_{-\mathbf{P}%
}^{+}a_{-\mathbf{P}}^{{}}+b_{-\mathbf{P}}^{+}b_{-\mathbf{P}}^{{}}) \ , \ \ \ \ \ \ \ \
\varepsilon_{mex}(P) = E_{band}(r)-
\mathcal{E}_{B}^{(b)} + \varepsilon _{0}(P) \ ,
\end{eqnarray}
%-------------------------------------------------------------------------
where $a_{\mathbf{P}}^{+}$, $a_{\mathbf{P}}^{{}}$, $b_{\mathbf{P}}^{+}$,
$b_{\mathbf{P}}^{{}}$ are creation and annihilation operators of the
magnetoexcitons with up and down dipoles. In Eq.~(\ref{h0}), $E_{band}(r)$ is the band gap  energy which can depend on the position of the magnetoexciton in space in the presence of the trap, $\mathcal{E}_{B}^{(b)}$ is the binding energy of a magnetoexciton, and $\varepsilon_{0}(P) = P^2/(2m_{B})$, where $m_{B}$ is the effective magnetic
mass of a magnetoexciton. Similarly to the  Bose atoms in a
trap in the case of a slowly varying
external potential \cite{Pitaevskii},   we can make the quasiclassical approximation,
 assuming that the effective magnetoexciton mass does not depend on a characteristic size $l$  of the trap and it is a constant within the trap.
  This quasiclassical approximation is valid if $P \gg \hbar/l$. The harmonic trap is formed by the two-dimensional planar potential in the plane of graphene.
  The potential trap can be produced in two
different ways. One way is when  the potential trap can be produced by
applying an external inhomogeneous electric field. The spatial
dependence of the external field potential $V(r)$ is caused by
shifting of  magnetoexciton energy by applying an external
inhomogeneous electric field. The photonic states in the cavity are
assumed to be unaffected by this electric field. In this case the
band energy  is given by $E_{band}(r) = E_{band}(0)+
V(r)$ ($E_{band}(0) =  \sqrt{2}\hbar v_{F}/r_{B}$ is the band
gap  energy, which is the difference between the Landau levels $1$ and $0$, $v_{F}$ is the Fermi velocity of electrons in graphene~\cite{Lukose}, $r_{B}=\sqrt{\hbar c/(eB)}$ is the magnetic length). Near the minimum of the magnetoexciton energy, $V(r)$ can be
approximated by the planar harmonic potential $\gamma r^{2}/2$,
where $\gamma$ is the spring constant, $r$ is the distance between the center of mass of
the   magnetoexciton and the center of the trap. Note that a high magnetic field
does not change the  trapping potential in the effective Hamiltonian \cite{Berman_L_S_C}.
Let us mention that, while the quasiparticles in GLs and QWs in high magnetic field are described by the same effective Hamiltonian with the only difference in the effective magnetic mass of magnetoexciton. This difference is caused by the four-component spinor structure of magnetoexciton wave function in GL, while magnetoexciton in a QW and CQWs is characterized by the  one-component scalar wave function.

The Hamiltonian which describes the interaction between magnetoexcitons is
%-------------------------------------------------------------------------
\begin{equation}
\begin{array}{c}
\label{hint}\hat{H}_{int}= (2S)^{-1}\sum_{\mathbf{P}_{1}+\mathbf{P}_{2}=%
\mathbf{P}_{3}+\mathbf{P}_{4}}^{{}}(U_{+}(a_{\mathbf{P}_{4}}^{+}a_{\mathbf{P}%
_{3}}^{+}a_{\mathbf{P}_{2}}a_{\mathbf{P}_{1}}+b_{\mathbf{P}_{4}}^{+}b_{%
\mathbf{P}_{3}}^{+}b_{\mathbf{P}_{2}}b_{\mathbf{P}_{1}})
-U_{-} a_{\mathbf{P}_{4}}^{+}b_{\mathbf{P}_{3}}^{+}a_{\mathbf{P}_{2}}b_{\mathbf{P}_{1}}),%
\end{array}%
\end{equation}%
%-------------------------------------------------------------------------b
where $U_{+}$ and $U_{-}$ are the 2D Fourier images of $U_{+}(R)$
and $U_{-}(R)$, respectively,  and $S$ is the surface of
the system.

The Hamiltonian and the energy spectrum of non-interacting photons in a semiconductor
microcavity are given by~\cite{Pau}:
%-------------------------------------------------------------------------
\begin{eqnarray}
\label{sp_phot} \hat H_{ph} = \sum_{{\bf P}}  \varepsilon _{ph}(P)
\hat{a}_{{\bf P}}^{\dagger}\hat{a}_{{\bf P}}^{} \ , \ \ \ \ \ \ \
\varepsilon _{ph}(P) = (c/n)\sqrt{P^{2} +
\hbar^{2}\pi^{2}L_{C}^{-2}} \ ,
\end{eqnarray}
%-------------------------------------------------------------------------
 where $\hat{a}_{{\bf P}}^{\dagger}$ and $\hat{a}_{{\bf P}}$ are
photonic creation and annihilation Bose operators.
In Eq.~(\ref{sp_phot}), $L_{C}$ is the length of the cavity, $n =
\sqrt{\epsilon_{C}}$ is the effective refractive index and
$\epsilon_{C}$ is the dielectric constant of the cavity. We assume that the
length of the microcavity has the following form:
%-------------------------------------------------------------------------
\begin{eqnarray}
\label{lc} L_{C}(B) = \frac{\hbar\pi c}{n \left(E_{band}(r) -
\mathcal{E}_{B}^{(b)}\right)} \ ,
\end{eqnarray}
%-------------------------------------------------------------------------
corresponding to the resonance of the photonic and magnetoexcitonic branches at $P = 0$, i.e.  $\varepsilon_{mex}(0) = \varepsilon_{ph}(0)$. As it follows from the energy spectra in~(\ref{h0}) and~(\ref{sp_phot}), and Eqs.~(\ref{emsa}) and~(\ref{lc}),
 the length of the microcavity, corresponding to a
magnetoexciton-photon resonance, decreases with the increment of
the magnetic field as $B^{-1/2}$. The resonance between
magnetoexcitons and cavity photonic modes can be achieved either by
controlling the spectrum  of magnetoexcitons $\varepsilon_{ex}(P)$
by changing magnetic field $B$ or by choosing  the appropriate length  of the
microcavity $L_{C}$.

 Alternatively to the case considered above, now the trapping of magnetopolaritons is caused by the
inhomogeneous shape of the cavity when the length of the cavity is determined by Eq.~(\ref{lc}) with the term $\gamma r^{2}/2$ added to $E_{band}(0) - \mathcal{E}_{B}^{(b)}$,
where $r$ is the distance between the photon
 and the center of the trap. In case 2, the $\gamma$  is the curvature characterizing the shape of the cavity.
The Hamiltonian and the energy spectrum of the photons in this case are shown by
Eq.~(\ref{sp_phot}), and the length of the microcavity is given by
Eq.~(\ref{lc}). In this case, for  the photonic spectrum in the effective mass approximation is given by  substituting the slowly changing shape of the length of cavity depending on the term  $\gamma r^{2}/2$ into Eq.~(\ref{sp_phot}) representing the spectrum of the cavity photons.
This quasiclassical approximation is valid if $P \gg \hbar /l$,
where  $l = \left(\hbar/(m_{B}\omega_{0})\right)^{1/2}$ is the size
of the magnetoexciton cloud in an ideal magnetoexciton gas and
$\omega_{0} = \sqrt{\gamma/m_{B}}$. The Hamiltonian and energy spectrum of
magnetoexcitons in this case are given by~(\ref{h0}).

 The Hamiltonian of the magnetoexciton-photon coupling has the form (see Refs.~[\onlinecite{Hopfield,Agranovich1,Ciuti,Agranovich2}]):
 %-------------------------------------------------------------------------
\begin{eqnarray}
\label{Ham_exph} \hat{H}_{mex-ph} = {\hbar \Omega_{R}}\sum_{{\bf P}}
 \hat{a}_{{\bf P}}^{\dagger}\hat{b}_{{\bf P}}^{} + h.c. \ ,
\end{eqnarray}
%-------------------------------------------------------------------------

The projection of the electron-hole Hamiltonian in magnetic field for the
CQWs and GLs onto the lowest Landau level results
in the effective Hamiltonian (\ref{H_eff}) with renormalized mass and where
term related to the vector potential is missing. The magnetic field in the
effective Hamiltonian (\ref{H_eff}) enters in the renormalized mass of the
magnetoexciton $m_{B}$. Therefore, Hamiltonian for the spatially separated
electrons and holes in two-layer system for the CQWs and  for the bilayer
graphene can be reduced in high magnetic field to the effective Hamiltonian (%
\ref{H_eff}). Magnetic field $B$ is reflected by the effective Hamiltonian (%
\ref{H_eff}) only through the effective magnetic mass of a
magnetoexciton $m_{B}$ in the expression for $\varepsilon _{0}(P)$ in the
first term of $\hat{H}_{mex}$. The only difference in the effective
Hamiltonian (\ref{H_eff}) for the CQWs and bilayer graphene realizations of
two-layer systems is that $m_{B}$ for the bilayer graphene is four times
less than for the CQWs due to the four-component spinor structure of the
wavefunction of the relative motion for the isolated non-interacting
electron-hole pair in magnetic field \cite{Berman_L_G}.

Transitions between Landau levels due to the Coulomb electron-hole
attraction for the large electron-hole separation $D\gg r_{B}$ can be
neglected, if the following condition is valid: $E_{b}=e^{2}/(\epsilon
_{b}D)\ll \hbar \omega _{c}=\hbar eB(m_{e}+m_{h})/(2m_{e}m_{h}c)$ for the
QWs and $E_{b}=4e^{2}/(\epsilon D)\ll \hbar v_{F}/r_{B}$ for the GLs, where  $v_{F}$ is the Fermi velocity of
electrons~\cite{Lukose}, $E_{b}$ and $\omega _{c}$ are the magnetoexcitonic binding energy and the
cyclotron frequency, respectively. This corresponds to the high magnetic
field $B$, the large interlayer separation $D$ and large dielectric constant
of the insulator layer between the GLs.

%-------------------------------------------------------------------------
\section{Bose-Einstein condensation of  polaritons in graphene in a high magnetic field}
\label{polariton}

In this Section we consider trapped polaritons in a single graphene layer embedded into an optical microcavity in high magnetic field. When an undoped electron system in graphene in a magnetic field without an external electric field is in the ground state, half of the zeroth Landau level is filled with electrons, all Landau levels above the zeroth one are  empty, and all levels below  the zeroth one are filled with electrons. We suggest using the gate voltage to control the chemical potential in graphene by two ways: to shift it above the zeroth level so that it is between the zeroth and first Landau levels (the first case) or to shift the chemical potential below the zeroth level so that it is between the  first negative and zeroth Landau levels (the second case). In both cases, all Landau levels below the chemical potential are completely filled and all Landau levels above the chemical potential are completely empty. Based on the selection rules for optical transitions between the Landau levels in  single-layer graphene~\cite{Gusynin3}, in the first case, there are allowed transitions between the zeroth and the first Landau levels, while in the second case there are allowed transitions between the first negative  and zeroth Landau levels.
  Correspondingly, we consider magnetoexcitons formed in graphene by the electron on the first Landau level and the hole on the zeroth Landau level (the first case) or the electron on the zeroth Landau level and the hole on the Landau level $-1$ (the second case). Note that by appropriate gate potential we can also use  any other neighboring Landau levels $n$ and $n+1$.

 For the relatively high dielectric constant of the microcavity,   $\epsilon \gg e^{2}/(\hbar v_{F}) \approx 2$ the magnetoexciton energy in graphene can be
calculated by applying perturbation theory with respect to the strength of the
Coulomb electron-hole attraction analogously as it was done in~[\onlinecite{Lerner}] for 2D quantum wells in a high magnetic field with non-zero electron $m_{e}$
and hole $m_{h}$ masses. This approach
allows us to obtain the spectrum
  of an isolated magnetoexciton with the electron on the Landau level $1$ and the hole on the Landau level $0$ in a single graphene layer, and it will be exactly the same as for the magnetoexciton with the electron on the Landau level $0$ and the hole on the Landau level $-1$.
The characteristic Coulomb electron-hole attraction for the single
graphene layer is $e^{2}/(\epsilon r_{B})$. The
energy difference between the first and zeroth Landau levels in
graphene is  $\hbar v_{F}/r_{B}$.    For graphene, the perturbative approach with respect to the strength of the Coulomb electron-hole attraction  is valid when $e^{2}/(\epsilon r_{B}) \ll \hbar
v_{F}/r_{B}$ \cite{Lerner}.
  This condition can be fulfilled at all magnetic
fields $B$ if the dielectric constant of the surrounding media satisfies the condition $
e^{2}/(\epsilon \hbar v_{F}) \ll 1$. Therefore, we claim that the
energy difference between the first and zeroth Landau levels is
always greater than the characteristic Coulomb attraction between the
electron and the hole in the single graphene layer at any $B$ if $\epsilon \gg e^{2}/(\hbar v_{F}) \approx 2$. Thus, applying perturbation theory with respect to weak Coulomb electron-hole attraction in graphene embedded in the $\mathrm{GaAs}$ microcavity ($\epsilon = 12.9$) is more accurate than for graphene embedded in the $\mathrm{SiO}_{2}$ microcavity ($\epsilon = 4.5$).  However, the magnetoexcitons in graphene exist in high magnetic field. Therefore, we restrict ourselves by consideration of high magnetic fields.

Polaritons are linear superpositions of excitons and photons. In
high magnetic fields, when magnetoexcitons may exist, the polaritons
become linear superpositions of magnetoexcitons and photons. Let
us define the superpositions of magnetoexcitons and photons as
magnetopolaritons.  It is obvious that magnetopolaritons in graphene
are two-dimensional, since graphene is a 2D structure.
The Hamiltonian of magnetopolaritons in the strong magnetic field is
given by Eq.~(\ref{Ham_tot_pol}). It can be shown that the interaction between two direct 2D
magnetoexcitons in graphene with the electron on the Landau level $1$ and the hole on the Landau level $0$ can be
neglected in a strong magnetic field, in analogy to what is described in Ref.~[\onlinecite{Lerner}] for 2D magnetoexcitons in a quantum well. Thus, the Hamiltonian $H_{tot}$ (\ref{Ham_tot_pol}) does not include
     the term corresponding to the interaction between two direct
     magnetoexcitons in a single graphene layer.  So in high magnetic field  there is the BEC of the ideal magnetoexcitonic gas in  graphene. Therefore, in a single graphene layer is high mahnetic field we assume $H_{int} = 0$ in Eq.~(\ref{Htotb}).

The binding energy $\mathcal{E}_{B}^{(b)}$ and effective magnetic mass $m_{B}$ of a magnetoexciton in graphene obtained using the first order perturbation respect to the electron-hole Coulomb attraction similarly to the case of a single quantum well \cite{Lerner} are given by
%%%%%%%%%%%%%%%%%%%%%%%%%%%%%%%%%%%%%%%%%%%%%%%%%%%%%%%%%%%%%%%%%%%%%%%%%%%%%%%%%%%%%%%%%%%%%%%%%%%%%%%%%%%%%%%%%%%%%%%%%%%%%%%%%%%%%%%%%%%%%%%%%%%%%%%%%%%
\begin{eqnarray}\label{emsa}
\mathcal{E}_{B}^{(b)} =   \sqrt{\frac{\pi}{2}}  \frac{e^{2}}{\epsilon r_{B}} ,
\hspace{3cm} m_{B} = \frac{2^{7/2}\epsilon \hbar^{2}}{\sqrt{\pi}e^{2}r_{B}} \ .
\end{eqnarray}
%%%%%%%%%%%%%%%%%%%%%%%%%%%%%%%%%%%%%%%%%%%%%%%%%%%%%%%%%%%%%%%%%%%%%%%%%%%%%%%%%%%%%%%%%%%%%%%%%%%%%%%%%%%%%%%%%%%%%%%%%%%%%%%%%%%%%%%%%%%%%%%%%%%%%%%%%%%

We obtain the effective Hamiltonian of polaritons by applying the standard procedure~\cite{Hopfield,Agranovich1,Ciuti,Agranovich2}, when we diagonalize the Hamiltonian $\hat{H}_{tot}$~(\ref{Ham_tot_pol}) by using Bogoliubov transformations.
If we measure the energy relative to the $P=0$ lower magnetopolariton
energy $(c/n) \hbar \pi L_{C}^{-1}  - |\hbar \Omega_{R}|$, we obtain
the resulting  effective Hamiltonian for trapped magnetopolaritons
in graphene in a magnetic field. At small momenta $\alpha \ll 1$
($L_{C} = \hbar\pi c/n \left(E_{band} -
\mathcal{E}_{B}^{(b)}\right)^{-1}$) and weak confinement $\beta
 \ll 1$, this effective Hamiltonian is
 %-------------------------------------------------------------------------
\begin{eqnarray}
\label{Ham_eff} \hat H_{\rm eff}  =
\sum_{\mathbf{P}}\left(\frac{P^{2}}{2M_{\rm eff}(B)} + \frac{1}{2}
V(r) \right)\hat{p}_{\mathbf{P}}^{\dagger}\hat{p}_{\mathbf{P}} \ ,
\end{eqnarray}
%-------------------------------------------------------------------------
 and  the effective magnetic mass of a magnetopolariton is
given by
%-------------------------------------------------------------------------
\begin{eqnarray}
\label{Meff} M_{\rm eff}(B) = 2   \left(m_{B}^{-1} + \frac{c
L_{C}(B)}{n\hbar\pi}\right)^{-1} \ .
\end{eqnarray}
%-------------------------------------------------------------------------
 According to Eqs.~(\ref{Meff}) and~(\ref{emsa}), the effective
magnetopolariton mass $M_{\rm eff}$ increases with the increment of
the magnetic field as $B^{1/2}$.
 Let us emphasize that the resulting effective Hamiltonian for magnetopolaritons  in graphene in a magnetic field for the parabolic trap is given by
Eq.~(\ref{Ham_eff}) for both physical realizations of
confinement represented by case 1 and case 2.

Neglecting anharmonic terms for the magnetoexciton-photon coupling,  the Rabi splitting constant $\Omega_{R}$  can
be estimated quasiclassically as
%-------------------------------------------------------------------------
\begin{eqnarray}
\label{defrabi}
\left|\hbar \Omega_{R}\right|^{2} =  \left| \left\langle  f \left|\hat{H}_{int} \right| i \right\rangle  \right|^{2} \ ,  \ \ \ E_{ph0} = \left(\frac{8 \pi \hbar \omega}{\epsilon W} \right)^{1/2} \ ,
\ \ \  \hat{H}_{int}  = -  \frac{v_{F}e}{c} \vec{\hat{\sigma}}\cdot \vec{A}  =   \frac{v_{F}e}{i \omega} \vec{\hat{\sigma}}\cdot \vec{E}_{ph0} \ ,
\end{eqnarray}
%-------------------------------------------------------------------------
where $\vec{\hat{\sigma}} = (\hat{\sigma}_{x},\hat{\sigma}_{y})$, $\hat{\sigma}_{x}$ and $\hat{\sigma}_{y}$ are Pauli matrices, $\hat{H}_{int}$ is the Hamiltonian of the electron-photon interaction corresponding to the electron in graphene described by Dirac dispersion, $E_{ph0}$ is the electric field corresponding to a single cavity photon,
 $W$ is the volume of microcavity,   $\omega$ is the photon frequency. The initial $| i \rangle$ electron state corresponds to the completely filled  Landau level $0$ and completely empty  Landau level $1$. The final $| f \rangle$ electron state corresponds to creation of one magnetoexciton with the electron on the Landau level $1$ and the hole on the Landau level $0$.
The transition dipole moment corresponding to the process of creation of this magnetoexciton is given by $d_{12} = e r_{B}/4$.
Let us note that in Eq.~(\ref{defrabi}) the energy of photon  absorbed at the creation of the magnetoexciton is given by $\hbar \omega = \varepsilon_{1} - \varepsilon_{0}  = \sqrt{2} \hbar v_{F}/r_{B}$ (we assume that $\mathcal{E}_{B}^{(b)} \ll \varepsilon_{1} - \varepsilon_{0}$).
Substituting the photon energy  and the transition dipole moment from  into   Eq.~(\ref{defrabi}), we obtain the Rabi splitting corresponding to the creation of a magnetoexciton with the electron on the Landau level $1$ and the hole on the Landau level $0$ in graphene: $\hbar \Omega_{R} = 2 e \left( \pi \hbar v_F r_{B}/(\sqrt{2} \epsilon W) \right)^{1/2}$.

Thus, the Rabi splitting in graphene is related to the creation of the magnetoexciton, which decreases when the magnetic field increases and is proportional to $B^{-1/4}$. Therefore, the Rabi splitting in graphene can be controlled by the external magnetic field.
It is easy to show that the Rabi  splitting related to the creation of the magnetoexciton, the electron on the Landau level $0$ and the hole on the Landau level $-1$ will be exactly the same as for the magnetoexciton with the electron on the Landau level $1$ and the hole on the Landau level $0$.
Let us mention that dipole optical transitions from the Landau level $-1$ to the  Landau level $0$, as well as from  the Landau level $0$ to the  Landau level $1$, are allowed by the selection rules for optical transitions in single-layer graphene \cite{Gusynin3}.

\begin{figure}[t] %  figure placement: here, top, bottom, or page
   \centering
  \includegraphics[width=3.5in]{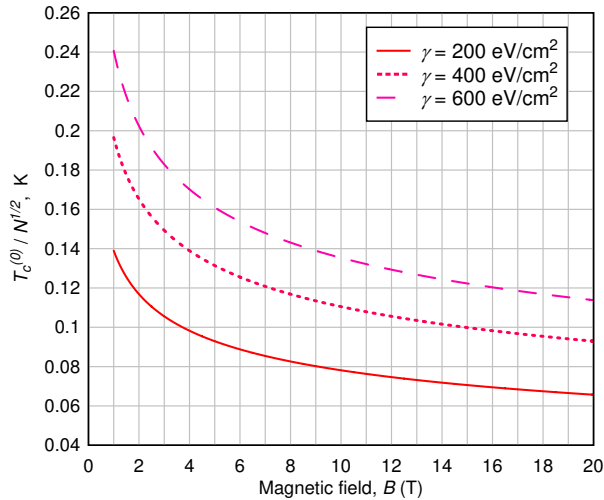}
   \caption{The ratio of the BEC critical temperature to the square root of the total number of magnetopolaritons $T_{c}^{(0)}/\sqrt{N}$ as a function of magnetic
   field $B$  at different spring constants $\gamma$. We assume the environment around graphene is $\mathrm{GaAs}$ with $\epsilon = 12.9$.}
   \label{fig_t_c}
\end{figure}

Although Bose-Einstein condensation cannot  take place in a 2D
homogeneous ideal gas at non-zero temperature, as discussed in
Ref.~[\onlinecite{Bagnato}], in a harmonic trap the BEC can occur in two
dimensions below a critical temperature $T_{c}^{0}$. In a harmonic
trap at a temperature $T$ below a critical temperature $T_{c}^{0}$ ($T
< T_{c}^{0}$), the number $N_{0}(T,B)$ of non-interacting magnetopolaritons in
the condensate  is given in Ref.~[\onlinecite{Bagnato}].
Applying the condition $N_{0}=0$, and assuming
that the magnetopolariton effective mass  is given by Eq.~(\ref{Meff}),
we obtain the BEC critical temperature $T_{c}^{(0)}$ for the ideal
gas of magnetopolaritons in a single graphene layer  in a magnetic
field:
%-------------------------------------------------------------------------
\begin{eqnarray}
 \label{t_c}
T_{c}^{(0)} (B) = \frac{1}{k_{B}}\left(\frac{3\hbar^{2}\gamma N}{\pi
 \left(g_{s}^{(e)}g_{v}^{(e)} +
 g_{s}^{(h)}g_{v}^{(h)}\right)M_{\rm eff}(B)}
\right)^{1/2} \ ,
\end{eqnarray}
%-------------------------------------------------------------------------
where $N$ is the total number of magnetopolaritons,
$g_{s}^{(e),(h)}$ and $g_{v}^{(e),(h)}$ are the spin and graphene
valley degeneracies for an electron and a hole, respectively,
$k_{B}$ is the Boltzmann constant.
At temperatures above $T_{c}^{(0)}$,  the BEC of magnetopolaritons in a single graphene layer does not exist.
$T_{c}^{(0)}/\sqrt{N}$ as a function of magnetic
   field $B$  and spring constant $\gamma$ is presented in
   Fig.~\ref{fig_t_c}. In our calculations, we used $g_{s}^{(e)} = g_{v}^{(e)} =
g_{s}^{(h)} = g_{v}^{(h)} = 2$.
 According to Eq.~(\ref{t_c}), the BEC critical temperature
$T_{c}^{(0)}$ decreases with the magnetic field  as
$B^{-1/4}$ and increases with  the spring constant
as $\gamma^{1/2}$. Note that we assume that the quality of the cavity is sufficiently high, so that the time of the relaxation to the Bose condensate quasiequilibrium state is smaller than the life time of the photons in the cavity.

Above we discussed the BEC of the magnetopolaritons in a single
graphene layer placed within a strong magnetic field. What would happen in  a
multilayer graphene system in a high magnetic field? Let us mention
that the magnetopolaritons formed by the microcavity photons and the
indirect excitons with the spatially separated electrons and holes
in different parallel graphene layers embedded in a semiconductor
microcavity can exist only at very low temperatures $k_{B} T \ll
\hbar \Omega_{R}$. For the case of the spatially separated electrons
and holes, the Rabi splitting $\Omega_{R}$ is very small in comparison to
the case of electrons and holes placed in a single graphene layer. This is
because $\Omega_{R} \sim d_{12}$ and the  matrix element of
magnetoexciton generation transition $d_{12}$ is proportional to the
overlapping integral of the electron and hole wavefunctions, which
is very small if the electrons and holes are placed in different
graphene layers. Therefore, we cannot predict the effect of
relatively high BEC critical temperature for the electrons and holes
placed in different graphene layers.

%-----------------------------------------------------------------------------------------------------------------------
%-----------------------------------------------------------------------------------------------------------------------

%-------------------------------------------------------------------------
\section{Superfluidity of magnetoexcitons in Bilayer Graphene}
\label{bilayer}

We consider two parallel graphene layers (GLs) separated by an insulating
slab of dielectric (for example, SiO$_2$) \cite{Berman_L_G}.  The spatial separation of electrons and holes in
different GLs can be achieved by applying an external
electric field. Besides, the spatially separated electrons and
holes can be created by varying the chemical potential by applying a
bias voltage between two GLs or between two gates
located near the corresponding graphene sheets.
The equilibrium system of local pairs of
spatially separated electrons and holes
 can be created by varying the chemical potential by using a
bias voltage between two GLs or between two gates
located near the corresponding graphene sheets (case 1) (for
simplicity, we also call these equilibrium local e-h pairs  in magnetic field $B$ as
indirect magnetoexcitons). In the case 1 a magnetoexciton is formed
by an electron on the Landau level $1$ and hole on the Landau level
$-1$. Magnetoexcitons with spatially separated electrons and holes
can be created also by laser pumping (far infrared in graphene) (case 2)
and by applying perpendicular electric field as for
CQWs\cite{Snoke,Butov,Eisenstein}. In the case 2 a magnetoexciton is
formed by an electron on the Landau level $1$ and hole on the Landau
level $0$.  We assume the system is in quasiequilibrium state.
 Below we assume the low-density regime for magnetoexcitons, i.e. magnetoexciton radius $a < n^{-1/2}$, where $n$ is the 2D magnetoexciton density.

In a strong magnetic field at low densities, $n\ll r_{B}^{-2}$ ($r_{B} = \left(\hbar c/(eB) \right)^{1/2}$ is the magnetic length,  $-e$ is the
electron charge, $c$ is the speed of light),
indirect magnetoexcitons repel as parallel dipoles, and we have for
the pair interaction potential:
%-------------------------------------------------------------------------
\begin{eqnarray}
\label{dipot} U(|\mathbf{R}_{1}-\mathbf{R}_{2}|) \simeq \frac{e^{2}D^{2}}{\epsilon |\mathbf{R}_{1}-%
\mathbf{R}_{2}|^{3}} \ ,
\end{eqnarray}
%-------------------------------------------------------------------------
where $D$ is the interlayer separation, $\epsilon$ is the dielectric constant of the insulator between two layers, $\mathbf{R}_{1(2)}$ are radius vectors of the center of mass of two magnetoexcitons. Since typically,
the value of $r$ is $r_{B}$, and $P\ll \hbar /r_{B}$ in this approximation,
the effective Hamiltonian $\hat{H}_{mex}$ in the magnetic momentum
representation $P$ in the subspace the lowest Landau level has the same
form (compare with Ref.[\onlinecite{Berman}]) as for two-dimensional
boson system without a magnetic field, but with the magnetoexciton
magnetic mass $m_{B}$ (which depends on $B$ and $D$; see below)
instead of the exciton mass ($M=m_{e}+m_{h}$), magnetic momenta
instead of ordinary momenta. We can obtain the effective Hamiltonian for bilayer graphene without a confinement if we keep considering only two graphene layers in the magnetoexciton effective Hamiltonian without a trap~(\ref{Htotb}):
%-------------------------------------------------------------------------
\begin{eqnarray}
\label{H_eff}
\hat{H}_{mex}=\sum_{\mathbf{P}}\varepsilon _{0}(P)\hat{a}_{\mathbf{P%
}}^{\dagger }\hat{a}_{\mathbf{P}}+\frac{1}{2}\sum_{\mathbf{P}_{1},\mathbf{P}%
_{2},\mathbf{P}_{3},\mathbf{P}_{4}}\left\langle \mathbf{P}_{1},\mathbf{P}%
_{2}\left\vert \hat{U}\right\vert \mathbf{P}_{3},\mathbf{P}_{4}\right\rangle
\hat{a}_{\mathbf{P}_{1}}^{\dagger }\hat{a}_{\mathbf{P}_{2}}^{\dagger }\hat{a}%
_{\mathbf{P}_{3}}\hat{a}_{\mathbf{P}_{4}},
\end{eqnarray}
%-------------------------------------------------------------------------
where the matrix element $\left\langle \mathbf{P}_{1},\mathbf{P}%
_{2}\left\vert \hat{U}\right\vert \mathbf{P}_{3},\mathbf{P}_{4}\right\rangle
$ is the Fourier transform of the pair interaction potential $%
U(R)=e^{2}D^{2}/\epsilon R^{3}$ and for the lowest Landau level we denote
the spectrum of the single exciton $\varepsilon _{0}(P)\equiv \varepsilon
_{00}(\mathbf{P})$. For an isolated magnetoexciton on the lowest Landau
level at the small magnetic momenta under consideration, $\varepsilon _{0}(%
\mathbf{P})\approx P^{2}/(2m_{B})$, where $m_{B}$ is the effective \textit{%
magnetic} mass of a magnetoexciton in the lowest Landau level and is a
function of the distance $D$ between $e$ -- and $h$ -- layers and magnetic
field $B$ (see Ref. [\onlinecite{Ruvinsky}]). In strong magnetic fields at $%
D\gg r_{B}$ the exciton magnetic mass is $m_{B}(D)=\epsilon
D^{3}/(e^{2}r_{B}^{4})$ for the QWs \cite{Ruvinsky} and $m_{B}(D)=\epsilon
D^{3}/(4e^{2}r_{B}^{4})$ for the GLs \cite{Berman_L_G}.

We study the magnetoexciton-magnetoexciton scattering applying the theory of weakly-interacting 2D Bose-gas~\cite{Lozovik,Berman}.
The chemical potential $\mu $ of two-dimensional dipole magnetoexcitons in graphene bilayer system, in the
 ladder approximation,   has the form
 (compare to Refs.~[\onlinecite{Lozovik,Berman}]):
%-------------------------------------------------------------------------
 \begin{eqnarray}\label{Mu}
\mu =  \frac{\kappa ^2 }{2m_{B}}
= \frac{\pi \hbar^{2}n}{sm_B \log \left[ s\hbar^{4}\epsilon^{2}/\left(2\pi n m_B^2 e^4 D^4\right) \right]} \ ,
\end{eqnarray}
%-------------------------------------------------------------------------
where $s=8$ is the spin and valley degeneracy factor for a magnetoexciton in graphene bilayer, $n$ is the 2D density of magnetoexcitons.

At small momenta the collective spectrum of magnetoexciton system
is the sound-like $\varepsilon (p) = c_{s}p$
  ($c_{s} = \sqrt{\mu/(2 m_{B})}$ is
the sound velocity) and  satisfied
to Landau criterium for superfluidity.
The density of the superfluid component
 $n_S (T)$  for two-dimensional system
with the sound spectrum can be estimated as:\cite{BLSC}
%-------------------------------------------------------------------------
\begin{eqnarray}
\label{n_s}
n_s =  n/(4s) -
 \frac{3 \zeta (3) }{2 \pi \hbar^{2}} \frac{k_{B}^{3}T^3}{c_s^4 m_B}\ .
\end{eqnarray}
%-------------------------------------------------------------------------
where $\zeta (z)$ is the Riemann zeta function, and $\zeta (3)
\simeq 1.202$. The second term in Eq.(\ref{n_s})  is the temperature
dependent
 normal density  taking into account  gas
of phonons ("bogolons") with dispersion law $\varepsilon (p) =
\sqrt{\mu/(2m_{B})}p$, $\mu $ is given by Eq.(\ref{Mu}).

\begin{figure}
\includegraphics[width = 3.5in]{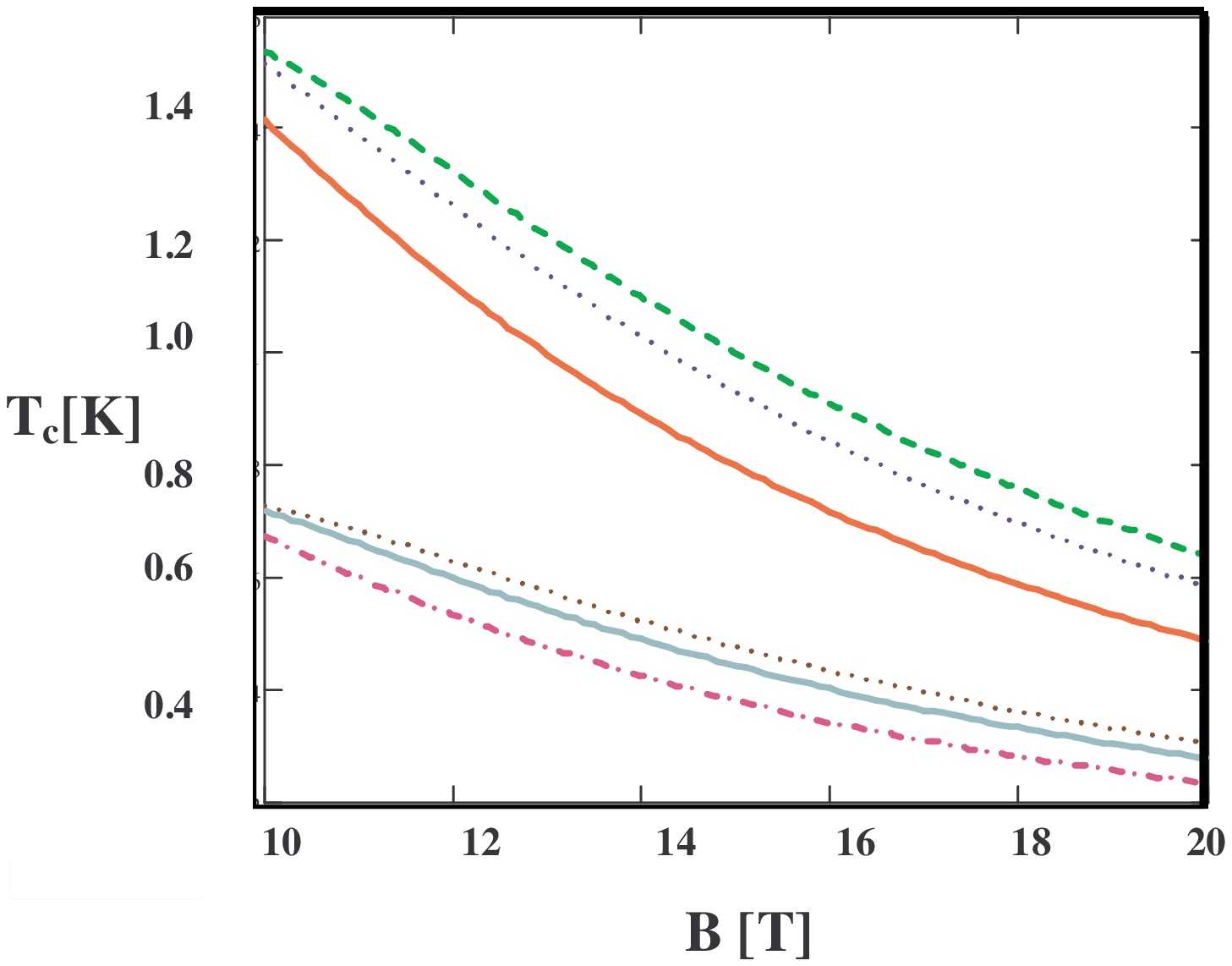}
\caption{ Dependence of  Kosterlitz-Thouless
transition temperature $T_c = T_c (B)$  (in units $K$) versus
magnetic field for two-layer graphene separated by SiO$_2$. with
$\epsilon_b = 4.5$. The magnetoexciton density $n = 4 \times
10^{11}cm^{-2}$. Different interlayer separations $D$ are chosen for
the case 1: $D = 30 nm$ (solid curve), $D = 28 nm$ (dotted curve),
$D = 27 nm$ (dashed curve). For the case 2: $D = 30 nm$
(dashed-dotted curve), $D = 28 nm$ (thin solid curve), $D = 27 nm$
(thin dotted curve). } \label{THD}
\end{figure}

In a 2D system, superfluidity of magnetoexcitons appears below the
Kosterlitz-Thouless transition temperature~\cite{Kosterlitz}: $T_{c} = \pi
n_{S}(T)/(2m_{B})$. The dependence of $T_c$ on $B$ and $D$  for the cases 1 and  2 is
represented in Fig.\ \ref{THD} (since in the case 1 the binding
energy two times higher and the effective magnetic mass is two times
smaller than in the case 2~\cite{Berman_L_G}, the magnetoexcitons in the case 1 are
expected to be twice more stable and $T_{c}$ in the case 1 is
expected to be approximately twice higher than in the case 2 at
fixed $n$, $D$ and $B$). The temperature $T_c$ for the onset of superfluidity due to the
Kosterlitz-Thouless transition at  a fixed magnetoexciton
density decreases as a function of magnetic field $B$ and
interlayer separation $D$. This is due to the increased $m_B$ as a
functions  of $B$ and $D$.  The $T_c$ decreases as $B^{-{1}/{2}}$ at
$D \ll r_B$ or as $B^{-2}$ when $D \gg r_B$.
%-------------------------------------------------------------------------
%-------------------------------------------------------------------------

%-------------------------------------------------------------------------
\section{Instability of dipole magnetoexcitons and superfluidity of magnetobiexcitons in graphene superlattices}
\label{super}

We consider magnetoexcitons in the
superlattices with  alternating electronic and hole  GLs. We suppose that recombination times can be much greater
 than relaxation times $\tau _{r}$
due to small  overlapping of spatially separation of electron and hole wave
functions in GLs. In this case electrons
and holes are characterized by different quasi-equilibrium chemical
potentials.
 Then  in the system of indirect excitons
in superlattices, as in CQW \cite{Lozovik,Berman}, the
quasiequilibrium phases appear. No external field applied to a slab of superlattice is assumed.  If
"electron" and "hole" quantum wells alternate, there are excitons
with parallel dipole moments in one pair of wells, but dipole
moments of excitons in another neighboring pairs of neighboring
wells have opposite direction. This fact leads to essential
distinction of properties of  $e-h$ system in superlattices from one
for coupled quantum wells with spatially separated electrons and
holes, where indirect exciton system is stable due to dipole-dipole
repelling of all excitons. This difference manifests itself already
beginning from three-layer $e-h-e$ or $h-e-h$ system. We assume that
alternating $e-h-e$ layers can be formed by independent gating with
the corresponding potentials which shift chemical potentials in
neighboring layers up and down or by alternating doping (by donors
and acceptors, respectively).

Let us show that the low-density system of weakly interacting
two-dimensional indirect magnetoexcitons in superlattices is
instable, contrary to the two-layer system in the CQW.
 At the small densities $n$  the system of indirect excitons at low temperatures
is the two-dimensional weakly nonideal Bose gas with normal to wells
dipole moments  $\mathbf{\it{d}}$ in the ground state ($d = eD$, $D$
is the interlayer separation). In contrast to ordinary excitons, for the
low-density spatially indirect magnetoexciton system the main
contribution to the energy is originated from the dipole-dipole
interactions $U_{-}$ and $U_{+}$ of magnetoexcitons with opposite
(see Fig.~\ref{biexciton}) and parallel dipoles, respectively.

\begin{figure}[tbp]
\includegraphics[width = 3.5in]{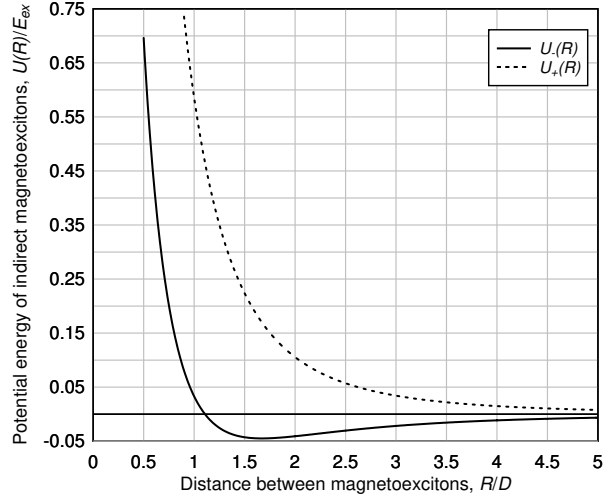}
\caption{The potential energy  of the interaction of indirect
magnetoexcitons with parallel $U_{+}(R)$ and  opposite $U_{-}(R)$
dipoles, located in neighboring pairs of  GLs (in units of the binding energy of the indirect magnetoexciton $%
E_{ex} = e^{2}/\protect\epsilon D$), as a function of the distances
$R$ between magnetoexcitons along the  GLs (in units of $D$).}
\label{dipoles}
\end{figure}

The behavior of the potential energies  $U_{+}(R)$ and $U_{-}(R)$ as
the functions of the distance between two excitons $R$ is shown in
Fig.~\ref{dipoles}.
 We suppose that $D/R\ll 1$ and $L/R\ll 1$, where $L$
is the mean distance between dipoles normal to the wells. We
consider the case, when the
number of quantum wells $k$ in superlattice is restricted $k\ll 1/(D\sqrt{%
\pi n})$. This is valid for small $k$ or for sufficiently low
exciton density.

 We can obtain the effective Hamiltonian for graphene superlattice without a confinement   if we keep considering only the magnetoexciton effective Hamiltonian~(\ref{Htotb}) without a trap. Let us apply the Bogolubov approximation to analyze a stability of
the ground state of the weakly nonideal Bose gas of indirect
excitons in superlattices.  We assume $U_{+}$ and $U_{-}$ are the 2D Fourier images of $U_{+}(R)$
and $U_{-}(R)$ at $P=0$, respectively,  and $S$ is the surface of
the system. Let us mention that the appropriate cut-off parameter
for this Fourier transform is the classical turning point of the
dipole-dipole interaction. Let us mention that the appropriate cut-off parameter
for this Fourier transform is the classical turning point of the
dipole-dipole interaction. Note that the  cut-off parameter $R_{0}$
for the potential $U_{+}(R)$ is much greater than for $U_{-}(R)$
(the cut-off parameters $R_{0}$ for the both potentials can be
represented in Fig.~\ref{dipoles} by the points where the curves
corresponding to $U_{+}(R)$ and $U_{-}(R)$ are crossed by the
chemical potential $\mu_{d}$ represented by the horizontal straight
line placed right above but close to $U_{+(-)}(R)=0$). Therefore, we
claim that $U_{+}>0$, $U_{-}<0$, and $|U_{-}| > |U_{+}|$.

Applying the unitary Bogoliubov transformations to the magnetoexciton operators $a_{\mathbf{P}}^{+}$, $a_{\mathbf{P}}^{{}}$, $b_{\mathbf{P}}^{+}$, and $b_{\mathbf{P}}^{{}}$, we diagonalize the Hamiltonian $\hat{H}_{tot}$ in the Bogoliubov approximation~\cite{Abrikosov}. Finally, we obtain
\begin{eqnarray}
\label{hdiag}
\hat{H}_{tot}=\sum_{\mathbf{P}\neq 0}^{{}}\varepsilon (p)(\alpha _{\mathbf{P}%
}^{+}\alpha _{\mathbf{P}}^{{}}+\beta _{\mathbf{P}}^{+}\beta _{\mathbf{P}%
}^{{}})
\end{eqnarray}
with the spectrum of quasiparticles $\varepsilon (P)$:
\begin{eqnarray}
\label{quasisp}
 \varepsilon_{1}^{2} (P)&=& \varepsilon_{0}^{2}(P)+ 2nU_{+}\varepsilon_{0}(P), \nonumber \\
\varepsilon_{2}^{2} (P) &=& \varepsilon_{0}^{2}(P)+ 2n(U_{+}+
U_{-})\varepsilon_{0}(P) \ .
\end{eqnarray}

 Since $U_{+} > 0$ and $U_{-} <0$, we have $\varepsilon_{1}^{2} (P)
> \varepsilon_{2}^{2} (P)$ at $P>0$. Therefore, at low  temperatures the quasiparticles only with
the spectrum $\varepsilon_{2}^{2} (P)$ will be excited, since the
excitation of these quasiparticles requires less energy than for the
quasiparticles with the spectrum $\varepsilon_{1}^{2} (P)$. Since
$U_{+} + U_{-} <0$, it is easy to see from Eq.~(\ref{quasisp}) that
for the small momenta $P< \sqrt{4m_{B}n|U_{+}+U_{-}|}$ the spectrum
of excitations becomes imaginary. Hence, the system of weakly
interacting indirect magnetoexcitons in the slab of the superlattice
is unstable. It can be seen that the condition of the instability of
magnetoexcitons becomes stronger as magnetic field higher, because
$m_{B}$ increases with the increment of the magnetic field, and,
therefore, the region of $P$ resulting in the imaginary collective
spectrum increases as $B$ increases.

As the ground state of the system  we consider the low-density
weakly nonideal gas of two-dimensional indirect magnetobiexcitons,
created by indirect magnetoexcitons with opposite dipoles
 in neighboring pairs of wells (Fig.~\ref{biexciton}).
The mean dipole moment of indirect
magnetobiexciton is equal to zero. However,
the quadrupole moment is nonzero and equal to $Q = 3eD^{2}$
(the large axis of the quadrupole is normal to quantum wells/graphene layers).
So indirect magnetobiexcitons interact at long distances $R \gg D$ as
parallel quadrupoles:  $U(R) = 9e^{2}D^{4}/(\epsilon R^{5})$.

We apply the theory of weakly-interacting 2D Bose gas~\cite{Lozovik} to study the magnetobiexciton-magnetobiexciton repulsion.
The chemical potential $\mu $ of two-dimensional biexcitons,
 repulsed by the quadrupole law, in the
 ladder approximation,   has the form
 (compare to Refs.~[\onlinecite{Lozovik,Berman}]):
 \begin{eqnarray}
\mu =  \frac{4\pi \hbar^{2}n_{bex}}{m_{B}^{b} \log \left[
\hbar^{4/3} \epsilon^{2/3}/\left(8\pi (18 m_B e^2
D^4)^{2/3}n_{bex}\right) \right]} . \label{Mu1}
\end{eqnarray}
 where $n_{bex} = n/8$
is the density of magnetobiexcitons in graphene layers and we considered that the magnetic mass of
magnetobiexciton is twice of the magnetic mass of magnetoexciton,
i.e. $2m_{B}$.

At small momenta the collective spectrum of magnetobiexciton system
is the sound-like $\varepsilon (p) = c_{s}p$
  ($c_{s} = \sqrt{\mu/(2 m_{B})}$ is
the sound velocity) and  satisfied
to Landau criterion for superfluidity.
The density of the superfluid component
 $n_S (T)$  for two-dimensional system
with the sound spectrum is~\cite{BLSC}
\begin{equation}
\label{n_s1} n_S(T) = n_{bex} - \frac{3 \zeta (3) }{4 \pi
\hbar^{2}}\frac{k_{B}^{3}T^3}{m_{B}c_s^4} \ .
\end{equation}

\begin{figure}[tbp]
\includegraphics[width = 4.2in]{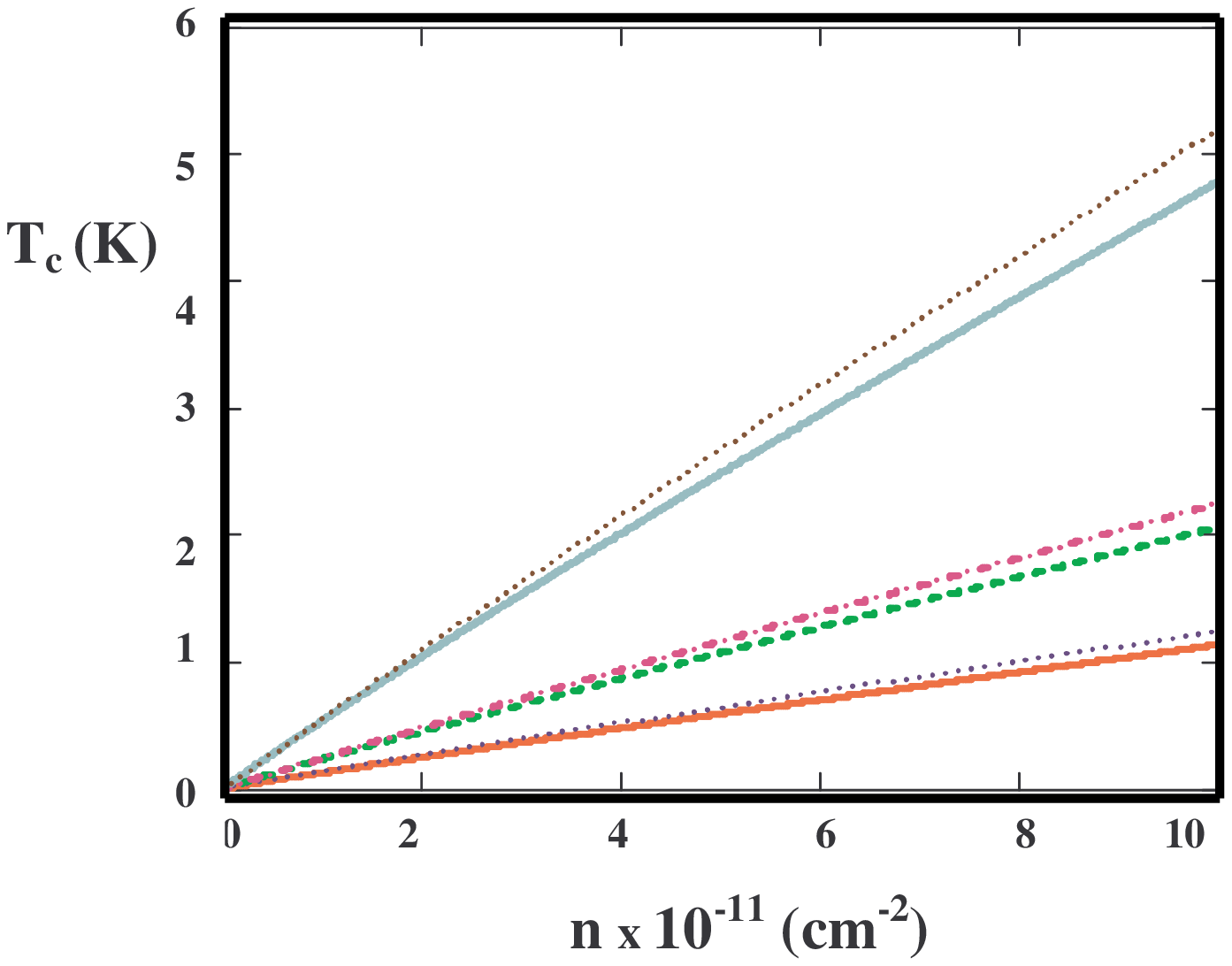}
\caption{Dependence of the Kosterlitz-Thouless transition
temperature $T_c =
T_c (B)$ for the superlattice consisting of the QWs for GaAs/AlGaAs, $%
\protect\epsilon = 13$; and for GLs separated by the layer of
SiO$_{2}$ with $\protect\epsilon = 4.5$ on the magnetoexciton
density $n$ at $D = 10$\ nm at different magnetic fields. The solid,
dashed and thin solid curves for the QWs, dotted, dashed-dotted and
thin dotted curves for the GL at $B$: $B = 20$\ T, $B = 15$\ T and
$B = 10$\ T, respectively.} \label{THD1}
\end{figure}

In a 2D system, superfluidity of magnetobiexcitons appears below the
Kosterlitz-Thouless transition temperature $T_{c} = \pi
n_{S}(T)/(4m_{B})$. Employing $n_S(T)$ for the superfluid component,
we obtain an equation for the Kosterlitz-Thouless transition
temperature $T_c$ $T_{c} = \pi
n_{S}(T)/(4m_{B})$. The dependence of $T_c$ on the density of magnetoexcitons at different
magnetic field $B$ for superlattice consisting of quantum wells and
graphene layers is represented on Fig.~\ref{THD1}. Let us mention, that we have apply the same effective Hamiltonian to describe magnetobiexcitons in high magnetic field in the superlattice of QWs and GLs with the only difference in the effective magnetic mass of magnetoexciton $m_{B}$.

According to  Fig.~\ref{THD1},  the temperature $T_c$ for the onset
of superfluidity due to the Kosterlitz-Thouless transition at  a
fixed magnetoexciton density decreases as a function of magnetic
field $B$ and interlayer separation $D$. This is due to the
increased effective magnetic mass $m_B$ of magnetoexcitons as a
functions of $B$ and $D$. The $T_c$ decreases as $B^{-{1}/{2}}$ at
$D \ll r_B$ or as $B^{-2}$ when $D \gg r_B$.
 According to Fig.~\ref{THD1}, the  Kosterlitz-Thouless temperature $T_{c}$ is higher for the superlattice consisting of  graphene layers than for the superlattice consisting of the quantum wells, and this difference is as stronger as the magnetic field is smaller.

\section{Conclusions}
\label{conc}

We have obtained the effective Hamiltonian of the quasiparticles in graphene structures in high magnetic field: indirect magnetoexcitons in graphene bilayer, magnetobiexcitons in graphene superlattices and magnetopolaritons in a graphene layer embedded in optical microcavity. It was shown that the gas of magnetoexcitons in graphene superlattice is instable due to the attraction between magnetoexcitons with parallel dipoles, while the system of magnetobiexcitons in the graphene superlattice is stable. We have shown that, while the quasiparticles in GLs and QWs in high magnetic field are described by the same effective Hamiltonian with the only difference in the effective magnetic mass of magnetoexciton. This difference is caused by the four-component spinor structure of magnetoexciton wave function in GL, while magnetoexciton in a QW and CQWs is characterized by the  one-component scalar wave function. Besides, we show that the magnetoexciton system graphene bilayer and magnetobiexciton system in graphene superlattice can be described as a 2D weakly-interacting Bose gas, which is superfluid below the Kosterlitz-Thouless phase transition temperature. We have calculated the density of the  superfluid component and the Kosterlitz-Thouless temperature for the systems of magnetoexcitons and magnetobiexcitons as functions of magnetic field $B$, interlayer separation $D$, and magnetoexciton density $n$. In contrast to the magnetoexcitons in graphene bilayer and magnetobiexcitons in graphene superlattice, the magnetopolaritons in GL embedded in an optical cavity in the limit of high magnetic field is ideal Bose gas without interparticle interactions, and, therefore, magnetopolariton gas is not superfluid. However, there is is BEC at the temperatures below critical one in this system in a trap. We have calculated the critical temperature of magnetopolariton BEC in GL embedded in an optical microcavity in a trap as a function of magnetic field and the curvature of the trap $\gamma$. Note that taking into account the virtual transitions of electrons and holes between Landau levels results in weak (at large $\epsilon$) interactions between magnetoexcitons~\cite{Lerner}. In turn, this leads to the possibility of the superfluidity of the magnetopolariton system.

%------------------------------------------------------------------------- %-------------------------------------------------------------------------

\end{document}